\documentclass[prl,twocolumn,showpacs]{revtex4}
\usepackage{amssymb}
\usepackage{graphicx}
\usepackage{dcolumn}
\usepackage{bm}

\begin{document}
\title{Suppression of Dephasing of Optically Trapped Atoms}
\author{M. F. Andersen, A. Kaplan, T. Gr\"{u}nzweig and N. Davidson.}
\affiliation{Department of Physics of Complex Systems,\\
Weizmann Institute of Science, Rehovot 76100, Israel}
\date{\today}

\begin{abstract}
Ultra-cold atoms trapped in an optical dipole trap and prepared in a coherent superposition of their hyperfine
ground states, decohere as they interact with their environment. We demonstrate than the loss in coherence in an
"echo" experiment, which is caused by mechanisms such as Rayleigh scattering, can be suppressed by the use of a
new pulse sequence. We also show that the coherence time is then limited by mixing to other vibrational levels in
the trap and by the finite lifetime of the internal quantum states of the atoms.
\end{abstract}

\maketitle

Long coherence time of trapped atoms have attracted interest over
the last years, not only because of their possible use in high
precision spectroscopy, but mainly because many proposed quantum
information processing schemes require manipulations of internal
states of atoms/ions, and any dephasing or decoherence of these
states will lead to loss of information. In the case of trapped
atoms or ions many different approaches have been tried for
increasing their coherence time. For trapped ions a ''dephasing
free subspace'' was used to minimize the dephasing induced by the
environment \cite{kielpinski01}. For neutral atoms and ions
several ''compensating'' techniques have been demonstrated, in
which the interaction causing the dephasing is canceled by an
additional interaction of opposite sign \cite {ido99}. Recently,
three groups have reported the use of coherence echoes (an analogy
to spin echoes \cite{Hahn50}) for investigating the coherence of
trapped alkali atoms and ions in a superposition state of two of
their ground states \cite{rowe02}\cite{andersen03}\cite{kuhr03}.
In \cite{andersen03} the increased coherence time was used to
study quantum dynamics of ultra cold atoms, and in \cite{rowe02}
and \cite{kuhr03} the study was related to quantum information
processing with single ions and atoms.

In this letter we investigate the limitations on the coherence
time achieved by echo techniques for ultra cold atoms trapped in
an optical dipole trap. A new echo technique for improving the
coherence time beyond the time achieved with the simple "$\pi
/2$-$\pi $-$\pi /2$" pulse sequence used in
\cite{rowe02},\cite{andersen03} and \cite{kuhr03} is demonstrated.
The improved coherence time is achieved by the use of additional
$\pi $-pulses between the two $\pi /2$-pulses. It relies on the
fact that dephasing is not an instantaneous effect, and if the
process is reversed before a complete dephasing has occurred, the
latter can be suppressed. The method therefore has a strong
parallel with the quantum Zeno effect \cite{itano90}, but we
stress that it is not a utilization of this effect, since in our
scheme no intermediate measurement is performed.

Any mechanism causing a change in the internal quantum state of an
atom, will clearly cause a drop in the ensemble coherence. On the
other hand, effects that lead to dephasing but conserve the
internal quantum state of the atom such as Rayleigh scattering of
far detuned photons \cite{cline94}, slow fluctuations of the trap
laser power and slow dynamics, will not cause an immediate
decoherence but will lead to a dephasing that grows in time. The
''multiple-$\pi $'' technique presented here can suppress
dephasing from these mechanisms.

We study the coherence of the two magnetic insensitive ground
states of $ ^{85}Rb$ atoms trapped in a far-off-resonance optical
dipole trap (FORT). These two levels ($\left|
5S_{1/2},F=2,m_{F}=0\right\rangle $, denoted $ \left|
1\right\rangle $, and $\left| 5S_{1/2},F=3,m_{F}=0\right\rangle $,
denoted $\left| 2\right\rangle $) are separated by the hyperfine
energy splitting $E_{HF}=\hbar \omega _{HF}$ with $\omega
_{HF}=2\pi \times 3.036$ GHz. Since the dipole potential is
inversely proportional to the trap laser detuning $\delta $
\cite{CCT91} there is a slightly different potential for atoms in
different hyperfine states $ \delta V=V_{2}\left( {\bf x}\right)
-V_{1}\left( {\bf x}\right) \simeq 2\times10^{-4}\times V_{1}$ for
our experimental parameters. This means that the entire
Hamiltonian can be written as: $H=H_{1}\left| 1\right\rangle
\left\langle 1\right| +H_{2}\left| 2\right\rangle \left\langle
2\right| =\left( \frac{p^{2}}{2m} +V_{1}\left( {\bf x}\right)
\right) \left| 1\right\rangle \left\langle 1\right| +\left(
\frac{p^{2}}{2m}+V_{2}\left( {\bf x}\right) +E_{HF}\right) \left|
2\right\rangle \left\langle 2\right| $, where $p$ is the atomic
center of mass momentum and $V_{1}$ [$V_{2}$] the external
potential for an atom in state $\left| 1\right\rangle $ [$\left|
2\right\rangle $], much smaller than $E_{HF}$. The atoms are
initially prepared in their internal ground state $\left|
1\right\rangle $. Their total wave function can be written as
$\Psi =\left| 1\right\rangle \otimes $ $\psi $, where $\psi $
represents the vibrational (external degree of freedom) part of
their wave function. A microwave (MW) field close to resonance
with $\omega _{HF}$ can drive transitions between the eigenstates
of the Hamiltonian corresponding to different internal states.
Since the radial size of our trap ($\sim 50$ $\mu m$) is much
smaller than the MW wavelength ($\sim $10 cm), the momentum of the
MW photon can be neglected \cite{CCT91}. The transition matrix
elements are thus given by $C_{nn^{\prime }}=\left\langle
n^{\prime }\mid n\right\rangle \times M_{1\rightarrow 2}$ where
$M_{1\rightarrow 2}$ is the free space matrix element for the
internal state transition, and $ \left\langle n^{\prime }\mid
n\right\rangle $ is the overlap between the initial vibrational
eigenstate of $H_{1}$ and an eigenstate of $H_{2}$ \cite{note4}.
In \cite {andersen03} it was shown that for a large detuning of
the FORT the potential difference $ \delta V $ is sufficiently
small to ensure $\left\langle n^{\prime }\mid n\right\rangle
\simeq \delta _{nn^{\prime }}$, for all populated $\left|
n\right\rangle $ of an atomic ensemble loaded into the FORT. Thus,
the atomic ensemble acts as an inhomogeneously broadened ensemble
of two-level systems, where the resonance frequency (energy
splitting) of each one depends on the vibrational eigenstate of
$H_{1}$ initially occupied by the atom. This effect is a direct
result of the quantization of the trap vibrational levels, and it
is strikingly evident even for a thermal ensemble with $k_{B}T\sim
10^{6}\times $mean level spacing (a condition that is often
considered enough to ensure validity of classical mechanics) as in
our experiment.

\begin{figure}[tb]
\begin{center}
\includegraphics[width=3in]{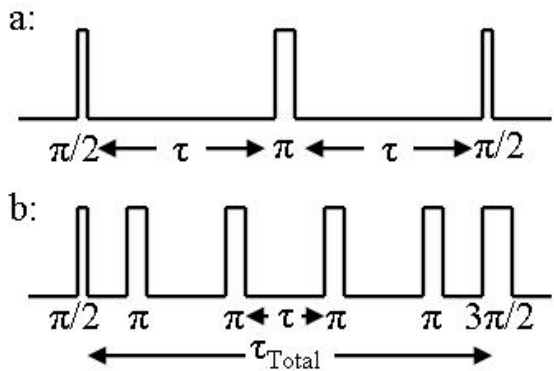}
\end{center}
\caption{a: Echo pulse sequence. b: multiple-$\pi$ pulse sequence. } \label{fipul}
\end{figure}

The echo pulse sequence consist of three pulses, two $\pi
/2$-pulses with a $ \pi $-pulse in between (see Fig.
\ref{fipul}a). The first $\pi /2$-pulse creates a coherent
superposition state of $\left| 1\right\rangle $ and $\left|
2\right\rangle $. After a time $\tau $ a $\pi $-pulse inverts the
populations, and after another time $\tau $ the atoms are exposed
to a second $\pi /2$-pulse, after which the interference of the
two parts of the wave function can be observed, as deviations of
the populations of states $\left| 1\right\rangle $ and $\left|
2\right\rangle $ from 1/2. For $\left\langle n\mid n^{\prime
}\right\rangle \simeq \delta _{nn^{\prime }}$, the above model
predicts that the echo pulse sequence will yield a perfect
interference, i.e. all the population will return to $\left|
1\right\rangle $ \cite{andersen03}. As was demonstrated in
\cite{kuhr03} this is not always the case, in real systems.

Two types of mechanisms can cause the echo coherence to decay for
long times. The first type of mechanisms (which can be denoted
"dynamical" $T_{2}$ processes) are processes leading to a time
dependent resonance frequency of the two-level system. This will
cause the two parts of the wave function generated by the first
$\pi /2$-pulse to acquire different phases during the two ''dark''
periods of time $\tau $, hence causing imperfect interference at
the time of the second $\pi /2$-pulse. Examples of such mechanisms
are existence of fluctuations of the trap depth e.g. due to noise
in the trap laser power \cite{kuhr03}, fluctuations in the bias
magnetic field giving rise to a fluctuating second order Zeeman
shift, collisions and spontaneous Rayleigh scattering of a photon
from the trap laser. Rayleigh scattering of photons does not lead
to instantaneous loss of coherence, as does a Raman scattering
process \cite{cline94}. Nevertheless, the recoil energy acquired
by the atom can significantly change its trap vibrational level,
and therefore its resonance frequency. The result is, again, a
time dependent resonance frequency of the two level systems. Other
heating mechanisms, such as pointing instability of the trap laser
beam, typically involve a much smaller energy change than Raleigh
photon scattering, hence they induce a much longer time scale for
dephasing.

The second type of mechanisms for decay of echo coherence ($
T_{1}$ processes) relates to the finite lifetime of the internal
states of the atoms, which is limited mainly due to transitions
induced by the trap laser light.

As stated above, dynamical $T_{2}$ processes do not cause
instantaneous decoherence, but require some time $\tau _{d}$ for a
substantial phase difference to evolve. For time between pulses
$\tau $ larger than $\tau _{d}$, we expect the coherent signal to
disappear. The time $\tau $ can be reduced by adding more (equally
spaced in time) $\pi $-pulses, between the two $\pi /2$ -pulses.
We expect a coherent signal to reappear for $\tau<\tau _{d}$, when
the decay of the echo coherence is dominated by dynamical $ T_{2}
$ processes.

In our experiment $^{85}Rb$ atoms are initially trapped in a
magneto optical trap, then cooled to a temperature of 20 $\mu K$
and loaded into the FORT by an optical molasses stage, that also
pumps the atoms into the F=2 hyperfine state. The FORT consists of
a 370 mW linearly polarized horizontal Gaussian laser beam focused
to a 1/e$^{2}$ radius of 50 $\mu m$ , and with a wavelength of
$\lambda $=810 $nm$ yielding a trap depth of $ U_{0}\approx 100$
$\mu K$. The power of the FORT is stabilized by a feedback loop to
a level of $\sim1\%$. The transverse oscillation time of atoms in
the trap is measured by parametric excitation spectroscopy to be
1.4 ms \cite{friebel98} ensuring $ \left\langle n^{\prime }\mid
n\right\rangle \simeq \delta _{n,n^{\prime }}$, for all thermally
populated transverse vibrational states \cite{andersen03}. The
free space Rabi-frequency of the atoms in the MW fields is 5 kHz.
A bias magnetic field of 250 mG (turned on after the atoms are
loaded into the trap) shifts all m$ _{F}\neq 0$ states out of
resonance with the MW field, limiting the MW transitions to the
two m$_{F}=0$ states ($\left| 1\right\rangle $ and $ \left|
2\right\rangle $). After the MW pulses the population of state $
\left| 2\right\rangle $ is detected by normalized selective
fluorescence detection \cite{andersen03}. We subtract from the
signal contributions to the population of $ \left| 2\right\rangle
$ due to F-changing Raman transitions induced by the trap laser
and normalize to the signal after a short $ \pi $-pulse.

\begin{figure}[tb]
\begin{center}
\includegraphics[width=3in]{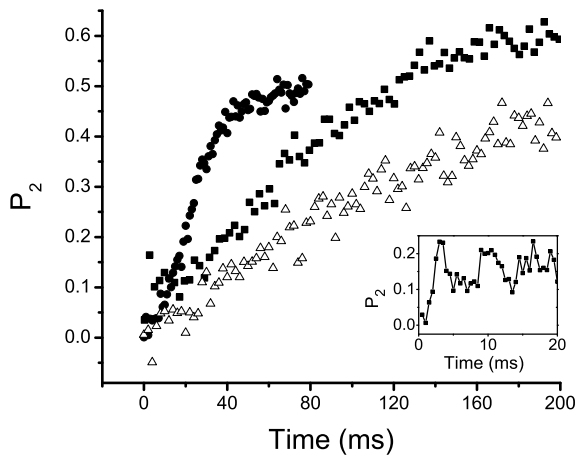}
\end{center}
\caption{$\bullet$ : Coherence signal ($P_{2}$) for the echo pulse sequence (Fig. 1a) as a function of $\tau
_{Total}$. $\blacksquare$ : $P_{2}$ for a pulse sequence with six $\pi$-pulses (Fig. 1b) as a function of $\tau
_{Total}$. It is seen that adding more $\pi$-pulses increases the coherence time, but also leads to an initial
small and rapid partial coherence decay. In the inset the short time signal for a pulse sequence with ten
$\pi$-pulses is shown. Wave-packet revivals are seen for $\tau_{Total}$=7 and 14 ms. $\vartriangle$ : $P_{2}$ for
a $\pi$-$\pi$ pulse sequence as a function of $\tau _{Total}$. A monotonic increase in the signal is seen, due to
transitions between m$_{F}$-states within the same F-manifold. } \label{fi1}
\end{figure}

As seen in Fig. \ref{fi1}, the echo signal starts from $P_{2}=0$
(indicating perfect coherence) and monotonicaly approaches
$P_{2}=\frac{1}{2}$ (indicating complete loss of coherence). The
coherence time $\tau _{c}$ is defined as the time $\tau _{Total}$
between the two $\pi /2$-pulses where $P_{2}$ reaches a value of
$P_{2}=\frac{1 }{2}(1-\frac{1}{e})$, and is seen to be $\tau
_{c}=$ 26 ms (the decay time for the fringes in a Ramsey
experiment is $\sim 5$ ms). We verify that this coherence time is
indeed not limited by longitudinal motion \cite{note4}, by
superimposing the trap with a $\lambda$=532 nm standing wave laser
field that confines the atoms in the longitudinal direction of the
trap and observing no improvement in the echo coherence time.

Next, we add more $\pi $-pulses using the pulse sequence shown in
Fig. \ref {fipul}b. First a $\pi /2$-pulse creates a coherent
superposition state of $\left| 1\right\rangle $ and $\left|
2\right\rangle $. After time $\tau /2$ the first $\pi $-pulse is
applied, followed by the rest of the $\pi $-pulses at time
intervals $\tau $. We end the sequence with a $3\pi /2$-pulse at
time $\tau /2$ after the last $\pi $-pulse in order to have
$P_{2}=0$ for a coherent signal for the even number of $ \pi
$-pulses that we use. The signal as a function of the total time
between the first and the last pulse, $\tau _{Total}$, for a pulse
sequence containing six $\pi $-pulses is also shown in Fig.
\ref{fi1}. It is seen that $\tau _{c}=65$ ms, clearly showing that
the additional $\pi $-pulses substantially increase the coherence
time.

\begin{figure}[tb]
\begin{center}
\includegraphics[width=3in]{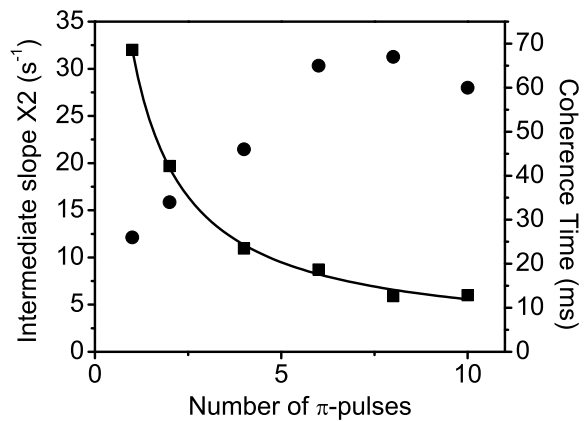}
\end{center}
\caption{$\bullet$ : Coherence time as a function of the number of $\pi$-pulses. A maximum coherence time of
$\sim65$ ms is observed, as a consequence of a trade-off between suppressing $T_{2}$-processes and the increased
mixing to different transverse vibrational states [5]. $\blacksquare$ : Twice the slope of the intermediate
regime, corrected for contributions from F-conserving transitions. Solid line: Fit to $\blacksquare$. The infinite
many $\pi$ decoherence rate is extrapolated from the fit, to yield a rate of 1.3 $s^{-1}$} \label{fi2}
\end{figure}

As shown in Fig. \ref{fi1}, adding more $\pi $-pulses leads to an
initial small and rapid partial loss of coherence. This is because
$\left\langle n^{\prime }\mid n\right\rangle $ is not strictly a
delta function, hence when more $\pi $-pulses are added, the
mixing to other transverse levels is increased and the dynamical
effects discussed in \cite{andersen03} appear (The asymptotic
value of $P_{2}$ due to this mixing effect is
$\frac{1}{2}(1-\left\langle n^{\prime }\mid
n\right\rangle^{2(n_{\pi}+1)}) $ where $n_{\pi}$ is the number of
$\pi $-pulses. The effect is too weak to be observed for
$n_{\pi}=1$). Shown in the inset, is the short time coherence
signal for a sequence with 10 $\pi$-pulses. Wave packet revivals
appear for time $\tau _{Total}=$7 and 14 ms (time between $\pi
$-pulses of 0.7 and 1.4 ms) as expected for a harmonic trap with
our measured transverse oscillation time of 1.4 ms
\cite{andersen03}.

In Fig. \ref{fi2} the coherence time is shown as a function of the
number of $\pi $-pulses. It is seen that the coherence time
initially shows linear dependence on the number of $\pi$-pulses
and then flattens out with a maximum value of $\sim65$ ms, a 2.5
fold improvement as compared to the simple echo coherence time.
The maximum coherence time is given by a trade-off between
suppressing the dephasing due to $T_{2}$ processes by adding more
$\pi$-pulses and the increased dephasing due to the mixing to
other transverse states, that the additional $\pi$-pulses induce.

The $P_{2}$ signal for a sequence with six $\pi$-pulses exceeds
1/2 at long times. We attribute this to a weak population mixing
between our "two-level" system, composed of levels $ \left|
1\right\rangle $ ($ \left| F=2, m_{F}=0\right\rangle $) and $
\left| 2\right\rangle $ ($ \left| F=3, m_{F}=0\right\rangle $) and
the m$_{F}\neq 0$ Zeeman states in the ground state hyperfine
manifold \cite{note2}. While our F-selective detection normalizes
and accounts for the F-changing transitions, it cannot account for
F-conserving, m$_{F}$-changing, transitions.

To isolate and directly measure the amount of coupling to other
m$_{F}$-states we measure the population of F=3 as a function of
time between two population-inverting $\pi $-pulses, where no
interference is involved. The results, also presented in Fig.
\ref{fi1}, clearly show that the population of F=3 increases as a
function of time between such two $\pi $-pulses, a fact that can
only result from coupling to the other m$ _{F} $-states
\cite{note3}. From the initial slope of this $\pi $-$\pi $
measurement the F-conserving transition rate is estimated as 1.2
$s^{-1}$ \cite{note1}. The F-changing transition rate was measured
independently to be 0.6 $s^{-1}$. Combining these two rates yields
a rate of 1.8 $s^{-1}$.

To investigate the ability of the multiple-$\pi$ scheme to
suppress dephasing due to dynamical $T_{2}$ processes, we are
interested in the slope of the intermediate decay (see Fig.
\ref{fi2}), found by fitting the intermediate section of the data
sets with a straight line, corrected for the long time slope,
thereby neglecting the effect of transverse motion and
F-conserving transitions (we show twice the slope since the decay
of coherence contribute only 1/2 to $P_{2}$). We see that the
addition of $\pi$-pulses improves the intermediate slope by a
factor of 6. To find the limiting value for the slope we assume
that dephasing from dynamical $T_{2}$ processes is suppressed
linearly with the number of $\pi$-pulses $n_{\pi}$. The data set
is then fitted with the function $a+b/(n_{\pi}-c)$, and the
limiting value $a$ is found to be $a$=1.3$\pm$0.8 $s^{-1}$
indicating that the above estimated total decay rate from the
m$_{F}$=0 states of 1.8 $s^{-1}$ plays a dominating role.

It should be noted that the geometrical shape of the trap
determines weather processes such as trap power fluctuations and
Rayleigh Scattering lead to dynamical $T_{2}$-processes. This can
be exploited to design a "dephasing free" trap, as proposed in
\cite{ande03b} in order to overcome the dephasing due to
inhomogeneous stark shift. In this trap also many of the above
listed processes will not lead to time dependent resonance
frequency of the two level systems.

In summary we demonstrated that the dephasing in MW spectroscopy
of optical trapped atoms due to dynamical $T_{2}$ processes can be
suppressed beyond the suppression offered by the simple echo, by
using an improved pulse sequence containing additional
$\pi$-pulses. The achieved coherence time is limited by increased
mixing between transverse states, and to a lesser extent the
lifetime of the internal states of the atoms. Both these factors
are expected to be substantially smaller for trap laser with much
larger detuning, such as in \cite{kuhr03}. This improvement is of
potential importance to the field of quantum information where
long coherence times are of great importance
\cite{rowe02}\cite{kuhr03}. Finally, the demonstrated pulse
sequence may also find use in precision spectroscopy of a periodic
effect, where the $\pi$-pulses can be synchronized with the period
of that effect.

\end{document}